\documentclass{elsart}

\def\Li#1#2{{\mathrm{Li}}_{#1}\left(#2\right)}
\def\ba{\begin{eqnarray}}
\def\ea{\end{eqnarray}}
\def\dd{{\mathrm d}}

\def\fun#1#2{\lower3.6pt\vbox{\baselineskip0pt\lineskip.9pt
  \ialign{$\mathsurround=0pt#1\hfil##\hfil$\crcr#2\crcr\sim\crcr}}}
\def\order#1{{\mathcal O}\left(#1\right)}

\begin{document}
\begin{frontmatter}

\title{First order radiative corrections to polarized muon decay spectrum}

\author{A.B. Arbuzov\thanksref{Turin}\thanksref{NSERC}}

\thanks[Turin]{The major part of this work was
performed in Dipartimento di Fisica Teorica, Universit\`a di Torino
\& INFN, Sezione di Torino, Italy.}

\thanks[NSERC]{This research was supported by the Natural Sciences 
and Engineering Research Council of Canada.}

\address{Department of Physics, University of Alberta\\
Edmonton, AB\ \  T6G 2J1, Canada\\
{\tt e-mail:} aarbuzov@phys.ualberta.ca
}

\begin{abstract}
The first order QED corrections to the polarized muon decay spectrum 
are considered. The exact dependence on electron and muon masses is kept. 
Numerical results are presented.
\end{abstract}


\end{frontmatter}

\section{Introduction}

Since the discovery of muon in 1936, experimental and theoretical
investigations of its properties became an important part of the 
elementary particle physics. The very accurate measurements of the properties
provide serious checks of the Standard Model and give a possibility
to look for new physics in low--energy experiments.
Because of the more and more precise experimental facilities and techniques,
calculations of radiative corrections become unavoidable. We need them to
obtain theoretical predictions with the required accuracy.

In this paper we will consider radiative corrections
to the polarized muon decay spectrum. This work was initialized by the 
experiment TWIST~\cite{Rodning:2001js,Quraan:2000vq}, which is going 
to measure the spectrum with
the error level of the order of $10^{-4}$. We will speak here only about
the first order quantum electrodynamic (QED) 
correction and concentrate on the effect of the non--zero
ratio of the electron and the muon masses. Higher order corrections to the
muon decay spectrum will be considered elsewhere~\cite{Davydychev:2001ee,mu_lla}.
A review about the muon properties and possible non--standard 
effects in the muon decay can be found in Refs.~\cite{Fetscher:2000th,Kuno:2001jp}.

\section{The Born level}

At the Born level within the Fermi Model the differential 
width is described by the formula~\cite{mich1,mich2}:
\ba
\frac{\dd^2\Gamma^{\mathrm{Born}}}{\dd x\dd c} 
&=& \Gamma_0
x^2\beta\left( f^{\mathrm{Born}}(x) + c\xi g^{\mathrm{Born}}(x) \right), \qquad
\Gamma_0 = \frac{G_F^2 m_\mu^5}{192\pi^3}\, ,
\nonumber \\
f^{\mathrm{Born}}(x) &=& 3 - 2x + \frac{x}{4}(3x-4)(1-\beta^2), 
\nonumber \\
g^{\mathrm{Born}}(x) &=& (1-2x)\beta + \frac{3x^2}{4}(1-\beta^2)\beta,
\nonumber \\
\beta &=& \sqrt{1-\frac{m_e^2}{E_e^2}}, \qquad E_e = \frac{m_\mu}{2}x, \qquad
c = \cos\theta,
\ea
where $m_e$ and $m_\mu$ are the electron and the muon masses;
$\theta$ is the angle between the electron momentum and the 
muon polarization vector ($c \to -c$ for the $\mu^+$ decay);
$\xi$ is the degree of the muon polarization;
$\beta$ is the electron velocity in the muon rest reference frame;
$E_e$ is the electron energy; $x$ is the electron energy fraction,
\ba
x_{\mathrm{min}} < x < x_{\mathrm{max}},\qquad
x_{\mathrm{min}} = 2\sqrt{\rho}, \qquad 
x_{\mathrm{max}} = 1 + \rho ,\qquad
\rho = \frac{m_e^2}{m_\mu^2}\, .
\ea
In the massless limit $(m_e \to 0)$ we have
\ba
f^{\mathrm{Born}}(x) \to f_0(x) = 3 - 2x, \qquad
g^{\mathrm{Born}}(x) \to g_0(x) = 1 - 2x.
\ea

The integration over the energy fraction gives
\ba
&& \int\limits_{x_{\mathrm{min}}}^{x_{\mathrm{max}}} \!\!\!\! \dd x \;
x^2\beta f^{\mathrm{Born}}(x) 
= \frac{1}{2}F(\rho), \qquad
\int\limits_{x_{\mathrm{min}}}^{x_{\mathrm{max}}} \!\!\!\! \dd x \;
x^2\beta g^{\mathrm{Born}}(x) 
= \frac{1}{2}G(\rho), 
\nonumber \\
&& F(\rho) =  1 - 8\rho - 12\rho^2\ln\rho + 8\rho^3 - \rho^4, 
\nonumber \\
&& G(\rho) = - \frac{1}{3} + \frac{32}{3}\rho^{3/2} - 30\rho^2
+ 32\rho^{5/2} - \frac{40}{3}\rho^3 + \rho^4.
\ea
Function $F(\rho)$ is relevant for the total decay width at the
Born level:
\ba
\Gamma^{\mathrm{Born}} = \int\limits_{-1}^{1}\dd c\; \Gamma_0 \frac{F(\rho)}{2}
= \Gamma_0 F(\rho).
\ea 
Function $G(\rho)$ contributes to the forward--backward asymmetry
of the decay:
\ba
\Gamma^{\mathrm{Born}}_{\mathrm{FB}} = \int\limits_{0}^{1}\dd c\; c\xi\; \Gamma_0 
\frac{G(\rho)}{2} - \int\limits_{-1}^{0}\dd c\; c\xi\; \Gamma_0 
\frac{G(\rho)}{2} = \xi \Gamma_0 G(\rho).
\ea

Within the Standard Model the muon decay
happens due to weak interaction of leptons and $W$-bosons. The Fermi
Model corresponds to the limiting case of the infinite $W$-boson mass.
We follow here the definition of the Fermi coupling constant $G_F$
as discussed in Ref.~\cite{vanRitbergen:2000fi}. That means, all weak
effects are incorporated into the coupling constant, and QED
radiative corrections
have to be calculated within the Fermi Model. 
I accept and support this approach.
Originally in the literature, the constant is defined in 
a different way~\cite{Groom:2000in,Marciano:1988vm,Sirlin:1980nh}, so that 
the first order effect in the muon and the $W$-boson mass ratio gives
\ba
\Gamma_0 \longrightarrow \Gamma_0 
\biggl( 1 + \frac{3}{5}\;\frac{m_\mu^2}{m_W^2} \biggr).
\ea
But in any case, in studies of the muon decay spectrum, it is natural 
to use the constant directly defined from very precise experiments
on the muon lifetime.
We have to note here, that although the Fermi lagrangian itself is not
renormalizable, QED corrections to the process under consideration
can be shown to be finite~\cite{Berman1962} at all orders of the
perturbation theory.

\section{The Exact First Order QED Corrections}

Here we will consider the $\order\alpha$ QED correction to the
muon decay spectrum with keeping the exact dependence on the electron
and the muon masses. The result is obtained by means of
the standard technique. The contributions from virtual, soft, and hard
photons were evaluated separately:
\ba
\frac{\dd^2\Gamma^{(1)}}{\dd x\dd c} =
\frac{\dd^2\Gamma^{\mathrm{Virt}}}{\dd x\dd c} + 
\frac{\dd^2\Gamma^{\mathrm{Soft}}}{\dd x\dd c} + 
\frac{\dd^2\Gamma^{\mathrm{Hard}}}{\dd x\dd c}\, . 
\ea

To be short, we give here only the simple formula for the soft 
photon contribution:
\ba
\frac{\dd^2\Gamma^{\mathrm{Soft}}}{\dd x\dd c} &=&
\frac{\dd^2\Gamma^{\mathrm{Born}}}{\dd x\dd c}\, \delta^{\mathrm{Soft}}, 
\\ \nonumber
\delta^{\mathrm{Soft}} &=& - \frac{\alpha}{2\pi}\biggl\{
2\biggl( 2\ln\frac{2\Delta\varepsilon}{m_\mu}
+ L + \ln\frac{m_e^2}{\lambda^2} \biggr)\biggl[ 1
- \frac{1}{2\beta}l_\beta \biggr] 
+ \frac{1}{2\beta}l^2_{\beta}
- \frac{1}{\beta}l_{\beta} 
\nonumber \\
&+& \frac{2}{\beta}\Li{2}{\frac{2\beta}{1+\beta}}
- 2 \biggr\}, \qquad
l_\beta = \ln\frac{1+\beta}{1-\beta}\; ,
\ea
where $\Delta\epsilon$ is the maximal energy of a soft photon
$(\Delta \ll 1)$;
$\lambda$ is a fictitious photon mass; functions $\Li{2}{x}$
and $\zeta(n)$ are defined in the Appendix.

The auxiliary parameters $\lambda$ and $\Delta\epsilon$
cancel out in the total sum  of the three contributions:
\ba
\frac{\dd^2\Gamma^{(1)}}{\dd x\dd c} 
&=& \Gamma_0\;x^2\beta\frac{\alpha}{2\pi}
\left( f_1(x) + c\xi g_1(x) \right),
\ea
\ba
f_1(x) &=& f^{\mathrm{Born}}(x)\biggl( \frac{2}{\beta}A
+ \frac{x^2(1-\beta^2)-4(1+x\beta)}{2x\beta}\ln\frac{q^2}{m_\mu^2} 
\nonumber \\
&+& \frac{4-x^2(1-\beta^2)}{x\beta}\ln\frac{2-x(1-\beta)}{2} \biggr)
\nonumber \\
&+& \frac{1}{\beta}\biggl( L + 2\ln x + 2\ln\frac{1+\beta}{2} \biggr)
\biggl\{
  \frac{5x^4}{384}(1-\beta^2)^3
- \frac{x^3}{4}(1-\beta^2)^2
\nonumber \\
&+& \frac{3x^2}{32}(3-12\beta+\beta^2)(1-\beta^2)
+ x\biggl[\frac{2}{3} + 2\beta
  + (1-\beta^2)\biggl( \frac{3}{2} + \beta \biggr)\biggr]
\nonumber \\
&+& \frac{1}{8}[ - 20 - 12\beta - 19(1-\beta^2)]
+ \frac{2}{x}
+ \frac{5}{6x^2}
\biggr\}
\nonumber \\
&+& \biggl(\ln x + \ln\frac{1+\beta}{2}\biggr)
\biggl[ \frac{9}{4}x^2(1-\beta^2) + 2x(\beta^2-3) + 3\biggr]
\nonumber \\
&+& f^{\mathrm{Born}}(x)\biggl[
- \frac{11}{18}x(1-\beta^2)
+ \frac{22}{27}\beta^2 - \frac{2}{9}
\biggr]
\nonumber \\
&+& x\biggl( - \frac{22}{27}\beta^4 + \frac{\beta^2}{2} 
 - \frac{11}{6} \biggr)
+ \frac{22}{9}( 3 - \beta^2 )
- \frac{22}{3x},
\ea

\ba
A &=& L \biggl( \ln\frac{q^2}{m_\mu^2} - \ln x
+ \ln\frac{1+\beta}{2\beta}
+ \ln\frac{2-x(1-\beta)}{2\beta} \biggr)
+ \biggl[ \ln\frac{q^2}{m_\mu^2} - 2\ln x
\nonumber \\
&+& 2\ln\frac{1+\beta}{2}
+ 4\ln\frac{2-x(1-\beta)}{2\beta}
\biggr]\biggl(\ln x + \ln\frac{1+\beta}{2}\biggr)
\nonumber \\
&+& 2\Li{2}{\frac{(1-\beta)(2-x(1+\beta))}{(1+\beta)(2-x(1-\beta))}}
- 2\Li{2}{\frac{2-x(1+\beta)}{2-x(1-\beta)}},
\ea

\ba
g_1(x) &=& g^{\mathrm{Born}}(x)\biggl( \frac{2}{\beta}A
- 4\ln\frac{2-x(1-\beta)}{2} \biggr)
\nonumber \\
&+& \frac{1}{\beta^2}\biggl( L + 2\ln x + 2\ln\frac{1+\beta}{2} \biggr)
\biggl\{
  \frac{5x^4}{384}(1-\beta^2)^3
\nonumber \\
&+& \frac{x^3}{8}(1-\beta^2)^2(1-3\beta^2)
+ \frac{3x^2}{32}(1-\beta^2)(-11+15\beta^2-12\beta^3)
\nonumber \\
&+& x\biggl[ \frac{2}{3} + 2\beta + (1-\beta^2)\biggl( \frac{\beta^2}{2}
  - 2\beta + \frac{3}{2}\biggr)\biggr]
- \frac{7}{2} - \frac{\beta}{2} 
\nonumber \\
&+& (1-\beta^2)\biggl(
  \frac{17}{8} + \frac{\beta}{2}\biggr)
- \frac{1}{6x^2}
\biggr\}
+ \beta\biggl(\ln x + \ln\frac{1+\beta}{2}\biggr)
\nonumber \\
&\times&\biggl( \frac{9}{4}x^2(1-\beta^2) - 4x + 1 \biggr)
+ \frac{1}{\beta^2}\biggl( \ln\frac{q^2}{m_\mu^2}
- 2\ln\frac{2-x(1-\beta)}{2}\biggr)
\nonumber \\
&\times&\biggl\{
- \frac{x^3}{48}(1-\beta^2)^2(1-19\beta^2)
+ x^2(1-\beta^2)\biggl( - \frac{3}{2}\beta^3
- \frac{5}{4}\beta^2 + \frac{1}{4} \biggr)
\nonumber \\
&+& x\biggl[ 4\beta 
+ (1-\beta^2)\biggl(- \frac{3}{4}\beta^2
- 4\beta - \frac{5}{4} \biggr)\biggr] 
+ \frac{16}{3} - 2\beta 
\nonumber \\
&+& (1-\beta^2)(2\beta - 2)
+ \frac{1}{x}(-6 + (1-\beta^2))
+ \frac{4}{x^2} - \frac{4}{3x^3}
\biggr\}
\nonumber \\
&+& g^{\mathrm{Born}}(x)\biggl[
- \frac{5}{144\beta^2}x^2(1-\beta^2)^2
- \frac{10}{27\beta^2}x(1-\beta^2)
- \frac{55}{54} + \frac{203}{162\beta^2} \biggr]
\nonumber \\
&+& \frac{x}{81}\biggl( \frac{17}{\beta} - 195\beta \biggr)
- \frac{1}{324}\biggl( \frac{595}{\beta} - 1923\beta \biggr)
+ \frac{10}{3x\beta} - \frac{1}{x^2\beta},
\nonumber \\
L &=& \ln\frac{m_\mu^2}{m_e^2}\, , \qquad
q^2 = m_{\mu}^2(1-x) + m_e^2.
\ea

Again the integration over the energy fraction and the angle 
gives us corrections to the decay width and the forward--backward asymmetry:
\ba \label{Gamma1}
\Gamma^{(1)} &=& \Gamma_0\;\frac{\alpha}{2\pi}F_1(\rho), \qquad
\Gamma^{(1)}_{\mathrm{FB}} = \xi\Gamma_0\;\frac{\alpha}{2\pi}G_1(\rho),
\ea
\ba \label{F1}
F_1(\rho) &=&  (1-\rho^2)\biggl(\frac{25}{4}
- \frac{239}{3}\rho + \frac{25}{4}\rho^2 \biggr)
- \rho\ln\rho \biggl( 20 + 90\rho - \frac{4}{3}\rho^2 
\nonumber \\
&+& \frac{17}{3}\rho^3\biggr) 
- \rho^2\ln^2\rho ( 36 + \rho^2 )
- (1-\rho^2)\biggl( \frac{17}{3} - \frac{64}{3}\rho 
+ \frac{17}{3}\rho^2\biggr)
\nonumber \\
&\times&\ln(1-\rho)  
+ 4( 1 + 30\rho^2 + \rho^4 )\ln\rho\ln(1-\rho)
+ 6( 1 + 16\rho^2 + \rho^4 )
\nonumber \\
&\times& [\Li{2}{\rho} - \zeta(2)]
+ 64\rho^{3/2}(1+\rho)\biggl(3\zeta(2) - 2\Li{2}{\sqrt{\rho}}
\nonumber \\
&+& 2\Li{2}{-\sqrt{\rho}} 
- \ln\rho\ln\frac{1-\sqrt{\rho}}{1+\sqrt{\rho}}
\biggr), 
\ea
\ba  \label{G1}
G_1(\rho) &=& - \frac{472}{27}(1-\sqrt{\rho})
+ (1-\rho)\biggl( \frac{1271}{108} + \frac{47}{27}\rho^{1/2}
- \frac{2959}{108}\rho + 60\rho^{3/2} 
\nonumber \\
&-& \frac{4657}{108}\rho^2 
+ 11\rho^{5/2} - \frac{21}{4}\rho^3 \biggr)
- \ln(1+\sqrt{\rho})(1-\rho)\biggl( \frac{100}{9} + \frac{52}{9}\rho
\nonumber \\
&+& \frac{268}{9}\rho^2 + \frac{20}{3}\rho^3 \biggr)
+ \ln(1-\sqrt{\rho})(1-\sqrt{\rho})\frac{2048}{9}
\nonumber \\
&+& \ln(1-\sqrt{\rho})(1-\rho)\biggl( - \frac{2035}{9}
+ \frac{2048}{9}\sqrt{\rho} 
- \frac{1987}{9}\rho + \frac{512}{3}\rho^{3/2} 
\nonumber \\
&-& \frac{637}{9}\rho^2 
+ \frac{17}{3}\rho^3 \biggr)
- \ln\rho(1-\sqrt{\rho})\frac{608}{9}
+ \ln\rho(1-\rho)\biggl( \frac{608}{9} 
\nonumber \\
&-& \frac{614}{9}\sqrt{\rho}
+ \frac{623}{9}\rho - 56\rho^{3/2} 
+ \frac{967}{18}\rho^2 + 2\rho^{5/2} 
+ \frac{1}{2}\rho^3 \biggr) 
\nonumber \\
&+& \ln\rho[ \ln(1-\sqrt{\rho}) 
+ 2\ln(1+\sqrt{\rho}) ]\biggl( \frac{2}{3} 
+ 8\rho^2 - \frac{32}{3}\rho^3 + 2\rho^4 \biggr)
\nonumber \\
&+& \rho^2\ln^2\rho\biggl( 7 + 16\rho - \frac{3}{2}\rho^2 \biggr)
+ ( \zeta(2) + 2\Li{2}{-\sqrt{\rho}} )\biggl( \frac{14}{3} + 16\rho 
\nonumber \\
&+& 32\rho^2 
- \frac{16}{3}\rho^3 + 6\rho^4 \biggr)
- \rho^2\frac{2}{3}\ln^3\rho
+ 16\rho^2\ln\rho[ 2\zeta(2) 
- 2\Li{2}{-\sqrt{\rho}} 
\nonumber \\
&-& \Li{2}{\sqrt{\rho}} ] 
+ 16\rho^2[ 2\Li{3}{\sqrt{\rho}} + 12\Li{3}{-\sqrt{\rho}} + 7\zeta(3) ].
\ea
The definitions of the special functions, used above, are given in
the Appendix.

The expression for $F_1(\rho)$ coincides with the one, received 
in Ref.~\cite{Nir:1989rm}, starting from a differential distribution of 
a different kind. We reproduced here the formula for the sake of 
completeness and for a comparison with $G_1(\rho)$.
Both the functions $F_1(\rho)$ and $G_1(\rho)$ are vanishing in the limit $\rho\to1$,
because of the vanishing phase space volume.

It is interesting to note, that $G_1(\rho)$ contains terms of the first power
in the electron and the muon mass ratio, while $F_1(\rho)$ does not 
(see the discussion about the odd mass terms in Ref.~\cite{vanRitbergen:2000fi}).
The linear mass terms in $G_1(\rho)$ can be clearly seen 
from the expansion\footnote{The expansion of $F_1(\rho)$
can be found in Ref.\cite{Nir:1989rm}.} of the exact formula:
\ba \label{G1exp}
G_1 &=& - \frac{617}{108} + \frac{14}{3}\zeta(2)
- \frac{8}{3}\sqrt{\rho}
+ \rho\biggl( - 32 + 16\zeta(2) + \frac{2}{3}\ln\rho \biggr)
\nonumber \\
&+& \rho^{3/2} \biggl( \frac{568}{27} + \frac{112}{9}\ln\rho \biggr)
+ \rho^2  \biggl( \frac{281}{6} + 32\zeta(2) + 112\zeta(3) - 16\ln\rho 
\nonumber \\
&+& 32\zeta(2)\ln\rho + 7\ln^2\rho - \frac{2}{3}\ln^3\rho \biggr)
+ \rho^{5/2}\biggl(  -\frac{95624}{225} + \frac{1232}{15}\ln\rho \biggr)
\nonumber \\
&+& \rho^3\biggl( \frac{5662}{27} - \frac{16}{3}\zeta(2) 
- \frac{698}{9}\ln\rho + 16\ln^2\rho \biggr)
\nonumber \\
&+& \rho^{7/2} \biggl(  - \frac{134248}{4725} - \frac{512}{63}\ln\rho \biggr)
+ {\mathcal O}(\rho^4).
\ea
One can check that the term with $|m_e/m_\mu|^1=\sqrt{\rho}$ 
is coming from the integration over the region of large (close to the
upper limit) values of the electron energy fraction. The absolute value
appears from the integration, it can be important for analytical
continuations.
Analogous terms with the first order mass ratio have been found in 
the forward--backward asymmetry in the process of electron--positron
annihilation into heavy leptons~\cite{Arbuzov:1992pr}.

The incorporation the first order correction gives the following
formula for the spectrum:
\ba
\frac{\dd^2\Gamma^{\mathrm{rad.corr.}}}{\dd x\dd c} = 
\frac{\dd^2\Gamma^{\mathrm{Born}}}{\dd x\dd c} +
\frac{\dd^2\Gamma^{(1)}}{\dd x\dd c} + {\mathcal O}(\alpha^2).
\ea

\section{Comparisons and numerical results}

The agreement with the massless formulae~\cite{Kinoshita:1959ru}
for $f_1(x)$ and $g_1(x)$ is checked. 
Our calculations for the massive case agree in part with the known results.
Namely, function $f_1(x)$ coincides with the result 
of Ref.~\cite{Behrends:1956mb}, where a mistake should be corrected
according to Ref.~\cite{Kinoshita:1959ru} (see also 
Appendix C in Ref.~\cite{vanRitbergen:2000fi}). 
The same function does
agree with the one given in Ref.~\cite{Marciano:1975kw}, while
the the integral over the real photon phase volume
has not been taken analytically there. Moreover, a confirmation
of the given formula for $f_1(x)$ comes from the comparison
of the result of its integration over the energy fraction
with Ref.~\cite{Nir:1989rm}, as mentioned above.

Function $g_1(x)$, which describes the polarized part of the decay spectrum,
could have been calculated long time ago on the same basis as $f_1(x)$.
Nevertheless, our results for $g_1(x)$ and $G_1(\rho)$ are new. 
A partial comparison between the virtual (loop) diagram contributions 
shows an agreement with the corresponding quantity from the calculation 
of the top--quark decay spectrum~\cite{Fischer:2001gp}~\footnote{Because 
of different choices of observables, the comparison of the total spectra 
is impossible.}. 

In Table~1 we show the effect of the mass terms at the lowest (Born) level
and at the level of the first order correction for different points of the
energy spectrum;
\ba
&& \delta_m^{\mathrm{Born}} = 10^4\cdot\biggl(
\frac{h^{\mathrm{Born}}(\tilde{x})}{h_0(x)} - 1\biggr), \qquad
\delta_m^{(1)} = 10^4\cdot\frac{\alpha}{2\pi}\;
\frac{h_1(\tilde{x})-h_1^{m_e\to 0}(x)}{h_0(x)}, 
\nonumber \\
&& h_{0,1}(x) = f_{0,1}(x)+c\xi g_{0,1}(x), \qquad 
h^{\mathrm{Born}}(\tilde{x}) = f^{\mathrm{Born}}(\tilde{x})
+ c\xi g^{\mathrm{Born}}(\tilde{x}), 
\ea
where the functions $f_1^{m_e\to0}(x)$ and $g_1^{m_e\to0}(x)$ 
can be received by applying $m_e=0$ in the general expressions
for $f_1(x)$ and $g_1(x)$ everywhere, except the argument of the
large logarithm $L$. In practice, the massless functions can be
taken from Ref.~\cite{Kinoshita:1959ru}. The argument of the
functions with the exact mass dependence is rescaled:
$\tilde{x} = x\cdot x_{\mathrm{max}}$.
\begin{table}
\caption{The effect of the finite electron--muon mass ratio versus the
electron energy fraction.}
\begin{tabular}[ht]{|r|r|r|r|r|r|r|}
\hline
\multicolumn{1}{|c|}{$x$} &
\multicolumn{1}{c|}{$h_0(x)$} & 
\multicolumn{1}{c|}{$h^{\mathrm{Born}}(\tilde{x})$} & 
\multicolumn{1}{c|}{$\delta_m^{\mathrm{Born}}$} & 
\multicolumn{1}{c|}{$h_1^{m_e\to0}(x)$} & 
\multicolumn{1}{c|}{$h_1(\tilde{x})$} & 
\multicolumn{1}{c|}{$\delta_m^{(1)}$} \\
\hline
\multicolumn{7}{|c|}{$c=0,\qquad \xi=1$} \\
\hline
 0.05 & 0.00725 & 0.00711 & $-$194.5 & 4.11481 & 4.10454 & $-$16.45 \\
 0.1 & 0.02800 & 0.02786 & $-$49.52 & 5.95508 & 5.95444 & $-$0.266 \\
 0.2 & 0.10400 & 0.10387 & $-$12.80 & 8.68399 & 8.68517 & 0.132 \\
 0.3 & 0.21600 & 0.21587 & $-$5.796 & 10.3054 & 10.3067 & 0.071 \\
 0.5 & 0.50000 & 0.49989 & $-$2.105 & 8.66761 & 8.66871 & 0.026 \\
 0.7 & 0.78400 & 0.78391 & $-$1.088 &$-$1.55489 &$-$1.55410 & 0.012 \\
 0.9 & 0.97200 & 0.97193 & $-$0.742 &$-$25.6678&$-$25.6670& 0.010 \\
0.99 & 0.99970 & 0.99963 & $-$0.702 &$-$67.6027&$-$67.6011& 0.019 \\
0.999& 1.00000 & 0.99993 & $-$0.702 &$-$107.665&$-$107.663 & 0.026 \\
\hline
\multicolumn{7}{|c|}{$c=1,\qquad \xi=1$} \\
\hline
0.05 & 0.00950 & 0.00928 & $-$236.8 &    3.6880 &    3.6657 & $-$27.32 \\
 0.1 & 0.03600 & 0.03579 & $-$59.02 &    5.2896 &    5.2850 & $-$1.490 \\
 0.2 & 0.12800 & 0.12781 & $-$14.51 &    7.4177 &    7.4179 &    0.027 \\
 0.3 & 0.25200 & 0.25184 & $-$6.186 &    7.8913 &    7.8925 &    0.054 \\
 0.5 & 0.50000 & 0.49991 & $-$1.871 &    2.7579 &    2.7589 &    0.024 \\
 0.7 & 0.58800 & 0.58796 & $-$0.659 & $-$8.1130 & $-$8.1129 &    0.001 \\
 0.9 & 0.32400 & 0.32400 & $-$0.150 & $-$13.120 & $-$13.121 & $-$0.022 \\
0.99 & 0.03920 & 0.03920 & $-$0.013 & $-$3.2805 & $-$3.2807 & $-$0.053 \\
0.999& 0.00399 & 0.00399 & $-$0.001 & $-$0.4949 & $-$0.4949 & $-$0.081 \\
\hline
\end{tabular}
\end{table}
One can see, that the effect of the mass terms in the ${\mathcal O}(\alpha)$
order is below the $10^{-4}$ precision tag (see $\delta_m^{(1)})$ for
experimentally~\cite{Rodning:2001js,Quraan:2000vq} preferable values of 
the electron energy fraction $x>0.3$.

In Table~2 we show the effect of the mass terms in the 
integrated quantities:
\ba
&& \delta_F^{\mathrm{Born}} = 10^4\cdot\biggl(\frac{F(\rho)}{F(0)} - 1\biggr), \qquad
\delta_F^{(1)} = 10^4\cdot\frac{\alpha}{2\pi}\;
\frac{F_1(\rho)-F_1(0)}{F(0)}, \nonumber \\
&& \delta_G^{\mathrm{Born}} = 10^4\cdot\biggl(\frac{G(\rho)}{G(0)} - 1\biggr), \qquad
\delta_G^{(1)} = 10^4\cdot\frac{\alpha}{2\pi}\;
\frac{G_1(\rho)-G_1(0)}{G(0)}\, .
\ea
\begin{table}
\caption{The effect of the finite mass ratio 
in the total decay width and in the forward--backward asymmetry.}
\begin{tabular}[ht]{|r|r|r|r|r|r|r|}
\hline
\multicolumn{1}{|c|}{$\rho$} &
\multicolumn{1}{c|}{$F(0)$} & 
\multicolumn{1}{c|}{$F(\rho)$} & 
\multicolumn{1}{c|}{$\delta_F^{\mathrm{Born}}$} & 
\multicolumn{1}{c|}{$F_1(0)$} & 
\multicolumn{1}{c|}{$F_1(\rho)$} & 
\multicolumn{1}{c|}{$\delta_F^{(1)}$} \\
\hline
$\frac{m_e^2}{m_\mu^2}=2.34\cdot10^{-5}$&
1.0000 & 0.9998 & $-$1.871 & $-$3.6196 & $-$3.6152 & 0.051 \\
$\frac{m_\mu^2}{m_\tau^2}=3.54\cdot10^{-3}$ &
1.0000 & 0.9726 & $-$274.3 & $-$3.6196 & $-$3.3367 & 3.286 \\
\hline
\multicolumn{1}{|c|}{$\rho$} &
\multicolumn{1}{c|}{$G(0)$} & 
\multicolumn{1}{c|}{$G(\rho)$} & 
\multicolumn{1}{c|}{$\delta_G^{\mathrm{Born}}$} & 
\multicolumn{1}{c|}{$G_1(0)$} & 
\multicolumn{1}{c|}{$G_1(\rho)$} & 
\multicolumn{1}{c|}{$\delta_G^{(1)}$} \\
\hline
$\frac{m_e^2}{m_\mu^2}=2.34\cdot10^{-5}$ &
$-$0.3333 & $-$0.3333 & $-$0.036 & 1.9634 & 1.9502 & 0.460 \\
$\frac{m_\mu^2}{m_\tau^2}=3.54\cdot10^{-3}$ &
$-$0.3333 & $-$0.3314 & $-$56.71 & 1.9634 & 1.7651 & 6.908 \\
\hline
\end{tabular}
\end{table}
The effect due to finite electron mass in  
function $G_1(\rho)$ is approaching the $10^{-4}$ level (for muon decay).

\section{Conclusions}

The present calculations can be easily extended for the case, where the
final electron polarization is measured. But, for the moment,
the experimental precision there does not call for an account
of small mass terms in the theoretical predictions.

In this way, we considered the first order QED radiative
correction to the muon decay spectrum. Formulae for the polarized part
of the spectrum $(g_1(x)$ and $G_1)$ with the exact dependence on the electron mass
are received for the first time.
The results presented are relevant for modern 
and future precise measurements of the muon decay spectrum. Our 
formulae are valid for the leptonic $\tau$ decays as well.

\ack{
I am grateful to A.~Czarnecki for fruitful discussions.
}

\section*{Appendix\\
Definition of special functions}

\setcounter{equation}{0}
\renewcommand{\theequation}{A.\arabic{equation}}

The Riemann $\zeta$-function and the polilogarithm functions are
defined as usually:
\ba
\zeta(n) &=& \sum_{k=1}^{\infty}\frac{1}{k^n}\, ,\qquad 
\zeta(2) = \frac{\pi^2}{6}\, ,\qquad 
\zeta(3) = 1.20205690315959\ldots 
\nonumber \\
\Li{2}{y} &=& - \int\limits_{0}^{y}\dd x\;\frac{\ln(1-x)}{x}\, ,\qquad
\Li{3}{y} = \int\limits_{0}^{y}\dd x\;\frac{\Li{2}{x}}{x}\, .
\ea
The following identities can help in numerical evaluations and estimates:
\ba
&& \Li{2}{1} = \zeta(2),\qquad \Li{2}{-1} = - \frac{1}{2}\zeta(2), \qquad
\Li{3}{1} = \zeta(3), \nonumber \\
&& \Li{3}{-1} = - \frac{3}{4}\zeta(3), \qquad
\Li{2}{y^2} = 2(\Li{2}{y} + \Li{2}{-y}),\nonumber \\
&& \Li{3}{y^2} = 4(\Li{3}{y} + \Li{3}{-y}).
\ea


\end{document}